\documentclass[english,aps,prl,amsmath,amssymb,twocolumn,letterpaper,showpacs,superscriptaddress]{revtex4-1}

\usepackage{mathtools} %floor and ceiling functions

\usepackage[normalem]{ulem} %Strikethrough

\usepackage{comment}
\usepackage{amsmath}
\usepackage{physics}
\usepackage{graphicx}
\usepackage{babel}
\usepackage{amstext}
\usepackage{esint}
\usepackage[unicode=true,pdfusetitle,
 bookmarks=true,bookmarksnumbered=false,bookmarksopen=false,
 breaklinks=false,pdfborder={0 0 1},backref=false,colorlinks=false]
 {hyperref}
\usepackage{pdfpages}	% Packages needed
\usepackage{pgffor}		% to include supplement

\makeatletter
\AtBeginDocument{\let\LS@rot\@undefined} %This fixes the issue
\makeatother							 %that pdfpages rotates the pages by 90deg.

\begin{document}

\title{Majorana signatures in charge transport through a topological superconducting double-island system}

\author{Jukka I. V\"ayrynen}

\affiliation{Department of Physics and Astronomy, Purdue University, West Lafayette, Indiana 47907 USA}

\affiliation{Station Q, Microsoft Corporation, Santa Barbara, California
93106-6105, USA}

\author{Dmitry I. Pikulin}

\affiliation{Microsoft Quantum, Redmond, Washington 98052, USA}

\affiliation{Station Q, Microsoft Corporation, Santa Barbara, California
93106-6105, USA}

\author{Roman M. Lutchyn}

\affiliation{Station Q, Microsoft Corporation, Santa Barbara, California
93106-6105, USA}

\affiliation{Quantum Science Center, Oak Ridge, TN 37830 USA}

\date{\today}
\begin{abstract}
We investigate the dynamics of a charge qubit consisting of two Coulomb-blockaded islands hosting Majorana zero modes. The frequency of the qubit is determined by coherent single-electron tunneling between two islands originating from the hybridization of two Majorana zero modes localized at the junction. We calculate the sequential tunneling current $I$  through the double-island device coupled to normal metal leads. We demonstrate that the $I\!-\!V$ characteristics in the large-bias regime can be used as a measurement of  the coherent Majorana coupling $E_M$. We propose dc and ac transport experiments for measuring $E_M$, and discuss their limitations due to {\it intrinsic} dephasing mechanisms such as excited quasiparticles. 
\end{abstract}
\maketitle

\begin{figure}
\centering
\includegraphics[width=0.95\columnwidth]{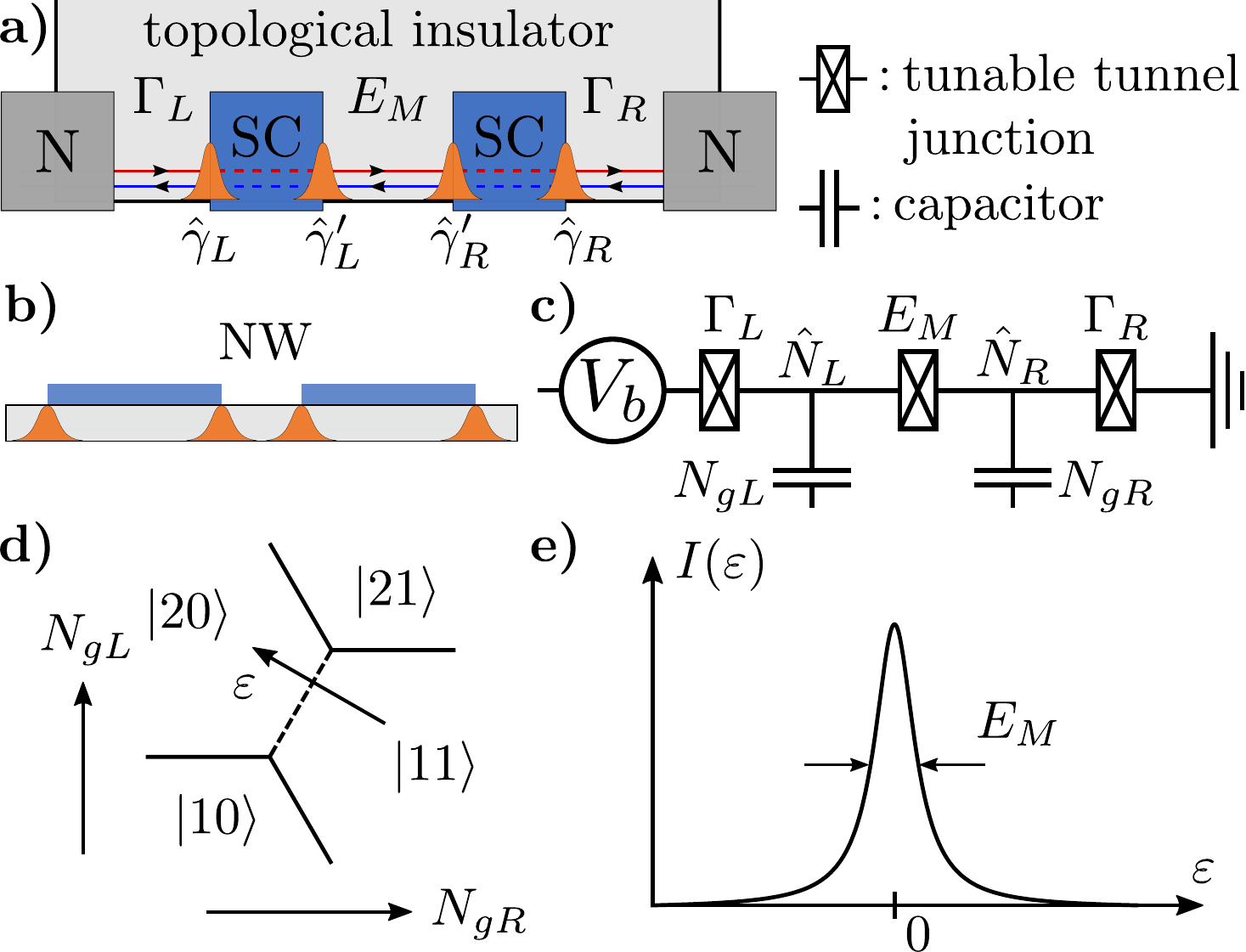}
\caption{\textbf{a--b)} Schematic of the topological superconductor (SC)  double island coupled to normal metal leads (N, grey). The double island can be  based on \textbf{a)} a 2D topological insulator edge~\cite{Fu09}, \textbf{b)} a semiconductor nanowire (NW)~\cite{2010PhRvL.105g7001L,*2010PhRvL.105q7002O,Lutchyn_REVIEW2018}, or a 3D topological insulator nanoribbon~\cite{
PhysRevB.84.201105,PhysRevB.98.205403} (not shown) and hosts four Majorana modes (orange). 
The outer Majoranas are incoherently coupled to the leads with rates $\Gamma_L,\Gamma_R$ while the inner two are hybridized with coupling $E_M$. 
These couplings are controllable by magnetic barriers (not shown) or depletion gates, depending on the platform.  
\textbf{c)} Circuit diagram corresponding to \textbf{(a--b)}. 
The double island charge  configuration $|N_L,N_R\rangle$ can be controlled with  dimensionless gate charges $N_{gL,R}$, see Eq.~(\ref{eq:Hc}).  
\textbf{d)} Charge stability diagram showing the ground state charge configurations for different gate-induced charges.  The lattice period  corresponds to $1e$, as opposed to $2e$ found in conventional superconducting islands~\cite{2016Natur.531..206A,van2020photon}.
The dashed line shows the position of a charge degeneracy ({\it i.e.}, $\epsilon=0$, see Eq.~\eqref{eq:epsilon}) which is lifted by $E_M$.  
\textbf{e)} Electrical current versus detuning from charge degeneracy point $\varepsilon$. When carried by the Majorana states, the current is a Lorentzian function of the detuning $\varepsilon$ with a width given by $E_M$.  This enables the measurement of the Majorana coupling $E_M$ in dc transport. } \label{fig:device}
\end{figure}

\textbf{\emph{Introduction.~}}
Topological qubits promise long lifetime and precise operations due to exponentially suppressed error rates.
The simplest topological qubit based on Majorana bound states %
requires four decoupled Majorana modes~\cite{Sarma_2015,Fidkowski2011,Vijay2016,2017NJPh...19a2001P,2017PhRvB..95w5305K, PhysRevX.6.031016,Manousakis17,PhysRevB.98.205403, pikulin2019quantum}. While demonstrating non-Abelian braiding statistics and developing a topological qubit 
are the current main objectives on the way towards  
topological quantum computation~\cite{Sarma_2015,Lutchyn_REVIEW2018}, it is useful to probe and demonstrate signatures of Majoranas in simple intermediate experiments. In this paper, we propose an experiment probing coherent coupling between two Majorana modes localized at a junction. The device, the \textit{Majorana charge qubit}, consists of two Coulomb-blockaded islands coupled together with a controllable Majorana coupling, see Fig.~\ref{fig:device}a.
Such a  double-island system  was recently used to demonstrate photon-assisted  tunneling of single electrons~\cite{van2020photon}. 

Our idea is similar to the one used to demonstrate coherent Cooper-pair tunneling in the first generation of the conventional superconducting charge qubits~\cite{nakamura1997spectroscopy,nakamura1999coherent}. Similarly, we propose 
to measure signatures of the coherent single-electron oscillations in transport through the device.
In the conventional superconducting case, the transport experiment
was made possible 
by the understanding of the so-called Josephson quasiparticle cycle~\cite{averin1989resonance} which was shown to give rise to  a resonant Josephson current  across the double island. 
The effect allows one to extract Josephson coupling $E_J$ between the two islands in a dc transport experiment.  
In this paper we focus on the analogous resonant current in a junction between two topological superconducting islands, see Fig.~\ref{fig:device}a. 
The tunable coupling $E_M$ between the two islands allows single electrons to tunnel coherently between the two Majorana zero modes (MZMs). 
Similar to the conventional case, the width of the current resonance can be used to estimate the hybridization energy $E_M$ between two MZMs, see Fig.~\ref{fig:device}e. 
We also propose a time-domain experiment involving  measurements of coherent $1e$-charge oscillations. 
We believe aforementioned experiments provide a simple way to detect single-electron coherent oscillations in a Majorana charge qubit via transport.   
In our proposal, the Majorana hybridization $E_M$ largely determines the frequency of the Majorana charge qubit. 
It is also a crucial component in  other Majorana-based proposals for quantum computing~\cite{%2017NJPh...19a2001P,2017PhRvB..95w5305K, PhysRevX.6.031016,Manousakis17,
2014NatCo...5E4772G,PhysRevB.98.205403,van2020photon}.

 As with superconducting qubits~\cite{nakamura1999coherent,lutchyn2006kinetics,glazman2020bogoliubov}, parasitic couplings to the environment lead to dephasing of a Majorana charge qubit~\cite{goldstein2011decay, rainis2012majorana, budich2012failure, 2015NatPh..11.1017H, Knapp_2018, karzig2021quasiparticle}. 
In this paper we focus on {\it intrinsic} dephasing mechanisms due to the presence of thermally activated or non-equilibrium quasiparticles (QPs).  
%w  
We  estimate  the influence of QPs on transport and how they may hinder the observation of the coherent $1e$ charge transfer. 
We obtain these estimates for Al-based devices where QP densities are well characterized. 
{\it Extrinsic} dephasing mechanisms are not well understood in nanowire and topological insulator (TI) systems; their effect can be incorporated into our formalism phenomenologically via parameter $\Gamma$. 

Here we focus on the sequential tunneling regime of large bias voltage $eV_b \gg E_C^{m}$, see Eq.~(\ref{eq:Hc}), in comparison to the mutual charging energy of the double island~\footnote{See Supplemental Material for details.}. We also assume that applied bias is smaller than the topological  superconducting gap, $eV_b \ll \Delta_P$, so that the transport is mostly carried by MZMs. The effect of above-gap quasiparticles is also included. We assume that most of the non-equilibrium QPs reside in the parent s-wave superconductor, whose gap we denote by $\Delta > \Delta_P$~\footnote{We assume for simplicity that there are no trivial subgap Andreev bound states; their contribution can be estimated by replacing $\Delta \to E_{ABS}$.}.  
In the charge qubit regime, we assume that the junctions are sufficiently closed for  the inter-island charging energy $E_C$, defined in Eq.~\eqref{eq:Hceff}, to be the dominant scale, $E_C \gg E_M,E_J$. % 

At low temperature, $T \ll E_C$, the current through the structure is given by a sequential-tunneling expression of the three junctions in series (see Fig.~\ref{fig:device}), 
\begin{equation}
  I(\varepsilon)= \frac{ e}{\gamma_{M}(\varepsilon)^{-1}+\Gamma_{L}^{-1}+\Gamma_{R}^{-1}}\,, \,\,\, \gamma_M (\varepsilon) =  \frac{ E_M^2 \Gamma_{\Sigma}}{ (\hbar\Gamma_{\Sigma})^{2}+4\varepsilon^{2}}\,, \label{eq:currentIdeal}
\end{equation}
where the rate  $\gamma_M (\varepsilon)$ across the central junction~\cite{averin1989resonance,nazarov2009quantum}  depends on the coherent coupling $E_M$ between the Majorana states, allowing its measurement from the current; $\Gamma_{L(R)}$ are the tunnel-couplings to outside left (right) normal metal lead and $\Gamma_{\Sigma}$ is the total dephasing rate; in the absence of other sources of dephasing the couplings to the leads provide the source of dephasing, $\Gamma_{\Sigma} \approx \Gamma_L +\Gamma_R$. Here we neglect effect of excited quasiparticles. 
The detuning $\varepsilon = 2E_C |\frac{1}{2} - \{N_g\}|$ is controllable with external gates ($\{\dots\}$ denotes the fractional part).

The current $I(\varepsilon)$ is a Lorentzian function of $\varepsilon$ with a width  $W = \max (\hbar\Gamma_{\Sigma},E_M \sqrt{ \Gamma_{\Sigma}(\Gamma_{L}^{-1}+\Gamma_{R}^{-1})} ) $. 
(In dimensionless gate voltage the width is $W/E_C$.) 
Assuming $W \ll E_C$, the current has a prominent  resonance at $\varepsilon =0$, see  Fig.~\ref{fig:device}e. 
The height of the Lorentzian $I(0)$ is given by the smallest of the rates $\Gamma_{L},\Gamma_{R}$, and $\gamma_{M}(0)=E_M^2/ (\hbar^2 \Gamma_{\Sigma})$.   
When $\gamma_{M}(0)$ is the largest rate, the current at resonance  is only weakly sensitive to $E_M$, but the width of the resonant peak is  $W \approx 2 E_M$ if we assume  $\Gamma_L \approx \Gamma_R$ and $\Gamma_{\Sigma} \approx 2 \Gamma_L$, see Fig.~\ref{fig:device}e. This allows one to estimate $E_M$ from the transport measurement.  
More generally $W \propto E_M$, with a prefactor that depends on $\Gamma/\Gamma_{L,R}$ where $\Gamma$ is the dephasing rate from other sources.  
While this might not be enough to extract the value of $E_M$, due to potentially unknown $\Gamma$,
one can use it to establish the functional dependence of $E_M$ on a controllable tunnel-barrier gate. 
When $\gamma_{M}(0)$ is the smallest rate, the height of the current peak is $I(0) = e E_M^2/ (\hbar^2 \Gamma_{\Sigma})$ and the width $W = \hbar \Gamma_{\Sigma}$. Under these conditions (ignoring QP transport)   the current is limited by the Majorana coupling, and one can thus extract $E_M$ from the product $I(0)W$.

The intra-island hybridization (splitting) of MZMs does not qualitatively change Eq.~(\ref{eq:currentIdeal}) but merely  shifts the resonance away from $\varepsilon =0$, see Eq.~(\ref{eq:Hceff}). 
This will also distort the charge stability diagram depicted in Fig.~\ref{fig:device}d, see Ref.~\cite{van2020photon}.

At finite temperature or in a non-equilibrium scenario, intrinsic dephasing  due to QP poisoning will modify the idealized results above.
Quasiparticles  open a parallel transport channel across the central junction which   introduces a new tunneling rate $\Gamma_{\text{qp}}$ and changes the form of current~(\ref{eq:currentIdeal}). %,
Similarly, the total dephasing  rate $\Gamma_{\Sigma}$ will increase due to $\Gamma_{\text{qp}}$; the full result for current is given in Eq.~(\ref{eq:I}). 
 %, 
Next, we introduce our model and derive the above results for the current. %
Throughout, we use units $\hbar = k_B = 1$.

\textbf{\emph{Model.~}} 
The charging Hamiltonian of the double island can be written as~\cite{Note1} %
\begin{equation}
H_{C}=E_{C}^{L}(\hat{N}_{L}-N_{gL})^{2}+E_{C}^{R}(\hat{N}_{R}-N_{gR})^{2}+E_{C}^{m}\hat{N}_{L} \hat{N}_{R}\,, \label{eq:Hc}
\end{equation}
where $\hat{N}_{L,R}$ are electron number operators of the left and right islands and $N_{gL,R} = - C_{gL,R} V_{gL,R}/e $ are dimensionless gate charges~\cite{Note1}, Fig.~\ref{fig:device}d. 
We assume that each island hosts a Majorana zero mode so that despite the islands being superconducting, both even and odd values of electron number are energetically allowed. 
We assume that each island is long enough so that we can ignore the intra-island hybridization between Majoranas.

We aim to probe the coherent coupling between the two Majoranas on different islands. This coupling conserves the total charge of the double-island but changes the charge distribution within it. For this reason, it is useful to project $H_C$ on a fixed total number of electrons $\hat{N}_L + \hat{N}_R$ and define an effective Hamiltonian that only depends on the charge difference $\hat{N}_{-} = (\hat{N}_L - \hat{N}_R)/2$. 
The effective charging Hamiltonian is~\cite{vdwRMP} 
\begin{equation}
H_{\text{eff}}=E_{C} (\hat{N}_{-}-N_{g})^{2}\,,\quad E_{C}=E_{C}^{R}+E_{C}^{L}-E_{C}^{m}\,, \label{eq:Hceff}
\end{equation}
where  $N_{g}=\frac{1}{E_{C}}\left(-E_{C}^{L}N_{gL}+E_{C}^{R}N_{gR}\right)$ up to a constant independent of $N_{gL}\,, N_{gR}$. 
Starting from a charge state $|N_{L},\,N_{R} \rangle$, the energy cost to transfer an electron from the left island MZM to the MZM on the right island is 
\begin{equation}
\varepsilon\equiv  U(N_{L}-1,\,N_{R}+1) - U(N_{L},\,N_{R}) =  2E_{C}(N_- + \frac{1}{2}-N_{g}) \,, \label{eq:epsilon}
\end{equation}
where we denote $U(N_{L},\,N_{R} ) =\langle N_{L},\,N_{R}  | H_C |N_{L},\,N_{R} \rangle$. The two states are degenerate, $\varepsilon=0$, when $N_g$ is half-integer. 
This is a charge degeneracy \textit{line} in the $N_{gL}$-$N_{gR}$ -plane, see Fig.~\ref{fig:device}b. 
We will focus on the vicinity of the charge degeneracy line $\epsilon =0$ where the current through the structure is the highest. 
The position of this degeneracy will shift if one of the MZM pairs is not at zero energy. 

The islands have neutral excitations   corresponding  to  broken Cooper pairs. 
We model the islands with a mean-field BCS Hamiltonian. 
After a Bogoliubov transformation~\cite{Note1} to  neutral quasiparticle operators $d_{n L,R}$, the neutral sector of the $\alpha = L,R$ island Hamiltonian becomes 
\begin{equation}
H_{\text{qp}}^{\alpha}=\sum_{n} \varepsilon_{n  \alpha} d_{n  \alpha}^{\dagger}d_{n  \alpha} \,, \label{eq:Hqp}
\end{equation}
where $\varepsilon_{n  \alpha}=\sqrt{\xi_{n, \alpha}^{2}+\Delta^{2}}$ and $\xi_{n, \alpha}$ are the normal state single-particle energies. We consider quasiparticles in the parent s-wave superconductor (Al) with gap $\Delta$. %
Each island also contains a fermionic level at zero energy [and therefore absent from Eq.~(\ref{eq:Hqp})] formed from a pair of decoupled Majoranas. 
The annihilation operator for an electron on the zero-energy level is $\hat{N}_{L,R}^{-} d_{L,R}$ with $d_{L,R} = (\gamma_{L,R}-i \gamma_{L,R}^{\prime})/2$ where $\gamma_{L,R},\gamma_{L,R}^{\prime}$ are neutral  Majorana operators. 
The operators $\hat{N}_{L,R}^{+ (-)}$ raise (lower)  the charge of the $L,R$ island, $\hat{N}_{L,R}^{+} = (\hat{N}_{L,R}^{-})^{\dagger}= |N_{L,R }+1 \rangle  \langle    N_{L,R} | $; they commute with the neutral operators $d_{n,\alpha}$, $d_{\alpha}$.

Tunneling of the  quasiparticles across the central junction dephases the Majorana charge qubit. 
The tunnel-coupling between the islands consists of quasiparticle tunneling and Majorana coupling, 
\begin{flalign}
H_{LR}  &=\sum_{nn'} t_{nn'}(d_{n L}^{\dagger} d_{n' R} u_{n}^{*} u_{n'} - d_{n'R}^{\dagger} d_{nL}   v_{n} v_{n'}^{*}) \hat{N}_{L}^{+} \hat{N}_{R}^{-} \nonumber \\
&+\frac{1}{2}E_{M}i\gamma_{L}^{\prime}\gamma_{R}^{\prime}\hat{N}_{L}^{+}\hat{N}_{R}^{-} +h.c.\,, \label{eq:HLR}
\end{flalign}
where we have ignored pair-breaking terms such as $d_{n L}^{\dagger} d_{n' R}^{\dagger}$ because they require a large energy transfer $\gtrsim\Delta$. This is a valid assumption in the voltage range  we are interested in, $eV_b \ll \Delta$.  
The first line in Eq.~(\ref{eq:HLR}) leads to incoherent inter-island tunneling rates $\Gamma_{\text{qp}}^{N_- \to N_- +1}$ 
and $\Gamma_{\text{qp}}^{ N_- +1 \to N_-  }$ which satisfy the detailed-balance condition  $\Gamma_{\text{qp}}^{ N_- \to N_- +1 }=e^{-\varepsilon/T}\Gamma_{\text{qp}}^{ N_- +1 \to N_-  }$.  
The second line in Eq.~(\ref{eq:HLR}) is the coherent Majorana coupling. 
(In a single-channel model~\cite{Kwon04,Fu09}, the coupling can be related to the topological gap and the  dimensionless conductance $g_c$ of the central junction:  $E_M =  \sqrt{g_c}\Delta_P$.)

In addition to neutral excitations, the double island  also has  excitations that change the total charge.  
For concreteness, we pick the degeneracy line between charge states $|2,\,0 \rangle$ and $|1,\, 1 \rangle$, see Fig.~\ref{fig:device}b. 
The lowest  charge excitations of the double island correspond to removing or adding an electron, with four relevant excitation energies 
$\varepsilon_{L}=U(2,0)-U(1,0)$,
$\varepsilon_{L}^{\prime}=U(2,1)-U(1,1)$, $\varepsilon_{R}=U(2,1)-U(2,0) $, and 
$ \varepsilon_{R}^{\prime}=U(1,1)-U(1,0)$. 
The coupling to outside normal metal leads generates these excitations. 
We describe the lead-island  couplings   by a tunneling Hamiltonian, 
\begin{equation}
H_t=\sum_{k}\left(t_{kL}\hat{N}_{L}^{+}\gamma_{L}c_{kL}+t_{kR}\hat{N}_{R}^{+}\gamma_{R}c_{kR}+h.c.\right)\,, \label{eq:Hlead-island}
\end{equation}
where $c_{kL,R}$ are the electron annihilation operators for the left and right leads, which are assumed to be spin polarized. 
Once again, we assume here relatively low bias voltage and ignore the tunneling from leads to above-gap quasiparticles in the island, see remarks below Eq.~(\ref{eq:HLR}). 
The Hamiltonian~(\ref{eq:Hlead-island}) yields incoherent tunneling rates ($\alpha=L,R$) ${\Gamma_{\alpha}  =\sum_{k}|t_{k\alpha}|^{2}2\pi\delta(E_{k\alpha}-\varepsilon_{\alpha}^{(\prime)})}$  between the leads and the islands. Here $ \varepsilon_{\alpha}^{(\prime)}$ is one of the four different energies required for removing/adding an electron from the double island. 
We   assume   uniform density of states in the leads so that the   rates $\Gamma_{L,R} $ are independent of the dot charge state. 
We also neglect  the $E_M$-induced modification to the  double-island spectrum, which is a valid approximation away from zero bias, $eV_b \gg E_M$.

\textbf{\emph{Rate equations.~}}
Let us next study the charge transport through the island. In the high-bias regime the dominant process is sequential tunneling rather than co-tunneling~\cite{nazarov2009quantum}. 
The voltage scale is set by the charging energy, $eV_b \gg E_C^m$~\cite{Note1}. 
For the sequential tunneling current, we can use a generalized rate equation where we include coherent inter-island coupling $E_M$ via the inner Majoranas~\cite{nazarov2009quantum}, see Eq.~(\ref{eq:HLR}). 
We assume that the islands are large enough so that the quasiparticle distribution relaxes to a steady state much faster than the charge distribution. This assumption is justified when quasiparticle energy relaxation rate is much faster than charge tunneling rates~\cite{PhysRevLett.103.097002,lutchyn2006kinetics}. 
To find charge distribution in steady state, we introduce a density matrix in the charge basis, $\hat{\rho}=\sum_{ij}p_{ji} \hat{P}_{i;j}$, where $i,j \in \{| N_L, N_R \rangle\}$ label the different charge states of the double-island and $\hat{P}_{i;j} =  | i \rangle \langle j|$ is a projector. 
We truncate the model to the 4 relevant charge states $| 1,0 \rangle $, $| 1,1 \rangle $, $| 2,0 \rangle $, and $| 2,1 \rangle $, see Fig.~\ref{fig:device}b. %
By using the Heisenberg equation of motion for $\hat{P}_{i;j}$, with Eqs.~(\ref{eq:Hc}), (\ref{eq:Hqp}), (\ref{eq:HLR}), and (\ref{eq:Hlead-island}) we find, 
\begin{widetext}
\begin{equation}
    \frac{d}{dt}\mathbf{p}=\boldsymbol{\Gamma}\mathbf{p}\,,\quad\boldsymbol{\Gamma}=\left(\begin{array}{cccccc}
-\Gamma_{\text{qp}}^{20\to11} & \Gamma_{\text{qp}}^{11\to20} & \Gamma_{L} & \Gamma_{R} & \frac{i}{2} E_{M} & -\frac{i}{2}E_{M}\\
\Gamma_{\text{qp}}^{20\to11} & -\Gamma_{\text{qp}}^{11\to20}-\Gamma_{L}-\Gamma_{R} & 0 & 0 & -\frac{i}{2}E_{M} & \frac{i}{2} E_{M}\\
0 & \Gamma_{R} & -\Gamma_{L} & 0 & 0 & 0\\
0 & \Gamma_{L} & 0 & -\Gamma_{R} & 0 & 0\\
\frac{i}{2} E_{M} & -\frac{i}{2} E_{M} & 0 & 0 & i\epsilon-\frac{1}{2}\Gamma_{\Sigma} & 0\\
-\frac{i}{2}E_{M} & \frac{i}{2}E_{M} & 0 & 0 & 0 & -i\epsilon -\frac{1}{2}\Gamma_{\Sigma}
\end{array}\right)\,, \label{eq:dpdt}
\end{equation}
where $\mathbf{p} = (p_{20;20},p_{11;11},p_{10;10},p_{21;21},p_{20;11},p_{20;11}^*)^T$. 
We have  replaced the lead distribution functions $n_{\alpha }(\epsilon_k) = \langle c_{k \alpha}^\dagger c_{k \alpha} \rangle$ by their 
high-bias 
values~\cite{Note1}. 
{
We denote  $\frac{1}{2}\Gamma_{\Sigma} =\frac{1}{2}T_1^{-1}+T_{\phi}^{-1}$ the total decoherence rate of the Majorana oscillations  in the equation for $dp_{20;11}/dt$. Here $T_1^{-1} = \Gamma_{\text{qp}}^{20\to11}+\Gamma_{\text{qp}}^{11\to 20}+\Gamma_{L}+\Gamma_{R}$ and $T_{\phi}^{-1} \equiv \Gamma$ correspond to energy relaxation rate and pure dephasing rate, respectively.
The latter rate may, for instance, include contributions due to extrinsic $1/f$ charge noise~\cite{Knapp_2018,van2020photon}. The estimate of $\Gamma$ depends on specific microscopic environment of the system and is beyond the scope of this work.}
\end{widetext}

\emph{ \textbf{Steady-state current.} }
The current in steady state  is equal to the rate of electrons flowing to the right lead~\cite{Note1}, $I = e \Gamma_R ( p_{11;11}  +  p_{21;21} )$, assuming high bias so that no electrons originate from the right lead, $n_R=0$. 
By using the steady state solution $d\mathbf{p}/dt=0$ from Eq.~(\ref{eq:dpdt}), 
we find, 
\begin{equation}
I=-e\frac{1}{\left[\gamma_{M}(\varepsilon)+\Gamma_{\text{qp}}^{\text{20\ensuremath{\to}11}}(\varepsilon)\right]^{-1}[1+\eta(\varepsilon)]+\Gamma_{L}^{-1}+\Gamma_{R}^{-1}}\,.
\label{eq:I}
\end{equation}
We have introduced the dimensionless parameter $\eta(\varepsilon) =  [\Gamma_{\text{qp}}^{\text{11\ensuremath{\to}20}}(\varepsilon)-\Gamma_{\text{qp}}^{\text{20\ensuremath{\to}11}}(\varepsilon)]/\left(\Gamma_{R}+\Gamma_{L}\right)$ which vanishes at $\varepsilon=0$. 
When quasiparticle tunneling is weak, $ \Gamma_{\text{qp}}^{\text{20\ensuremath{\to}11}}(0) \ll \gamma_{M} (0)$, 
the transport through the central junction is carried by the Majorana states  and current vs. detuning is  given in Eq.~(\ref{eq:currentIdeal}). In the considered high-bias limit, the finite-temperature corrections to Eq.~(\ref{eq:currentIdeal}) only appear at relatively high temperature $T \sim E_C$.  
On the other hand, if $E_M$ is too small, $ \Gamma_{\text{qp}}^{\text{20\ensuremath{\to}11}}(0) \gg \gamma_{M}(0) $, the quasiparticle contribution   dominates the transport through the middle junction in Eq.~(\ref{eq:I}). 
In this regime of quasiparticle dominated transport, the $I$ vs $\varepsilon$ is no longer Lorentzian but rather the current decays exponentially, $I(\varepsilon) \propto e^{-|\varepsilon|/2T}$, see  Eq.~(\ref{eq:Gammaqp}) below. 
In order to observe the resonant current broadened by $E_M$, Fig.~\ref{fig:device}c, the rate $\gamma_{M}(0)$ has to be the largest one. 
Since $\gamma_M(0) = E_M^2/\Gamma_{\Sigma}$, we see that for the device to be in the Majorana-dominated transport regime, $E_M$ needs to be larger than the scale $ \Gamma_{\Sigma}$, where  $\Gamma_{\Sigma}$ is approximately equal to the largest incoherent rate,  $\Gamma_{\Sigma} \approx  \max (\Gamma_{\text{qp}},\Gamma_{L,R}, \Gamma) $. 
In the next Section, we estimate the rates $\Gamma_{\text{qp}}^{\text{20\ensuremath{\to}11}}$ and $\Gamma_{L,R}$ to derive the lower bound for a measurable $E_M$.

\textbf{\emph{Estimation of the required regime of parameters.}} 
 In this Section, we connect the phenomenological rates in Eq.~(\ref{eq:I}) to the microscopic Hamiltonians, Eqs.~(\ref{eq:Hc}) and (\ref{eq:Hqp})--(\ref{eq:Hlead-island}). The rates $\Gamma_L,\Gamma_R$ of lead-island tunneling into the Majorana state can be estimated from Eq.~(\ref{eq:Hlead-island}) once $t_{k \alpha}$ is specified. 
This calculation has been carried out in  Ref.~\cite{PhysRevB.93.235431} and one finds
\begin{equation}
\Gamma_{\alpha} \! =2\pi\sum_{k}|t_{k\alpha}|^{2}\delta(\varepsilon_{k\alpha}-\varepsilon_{\alpha})=\frac{1}{\pi}g_{\alpha}\Delta\,, \,\, \alpha = L,R , \label{eq:GammaLR}
\end{equation}
where $g_{\alpha}=G_{\alpha}/(e^{2}/h)$ is the normal state dimensionless conductance of
the junction to lead $\alpha$. 
Since the junction conductances are independently controllable, it should be straightforward to enter the regime $\Gamma_L,\Gamma_R \ll E_M$.

For the QP tunneling rate, we use Eqs.~(\ref{eq:Hqp})--(\ref{eq:HLR}) to find~\cite{catelani2011}  
\begin{equation}
\Gamma_{\text{qp}}^{ 20\to 11 } (\varepsilon)=\frac{1}{2\pi}g_{c}  \sqrt{\frac{2\Delta}{|\varepsilon| }}  \nu_{0}^{-1} n_{\text{qp}} 
e^{-|\varepsilon|/2T}\,, \quad  T\ll\varepsilon\ll\Delta \,,
\label{eq:Gammaqp}
\end{equation}
where we denote $\pi|t|^{2}\nu_{0}^{2}=\frac{1}{2\pi}g_{c}$. Here $g_{c}$
is the normal-state dimensionless conductance of the central
junction and $\nu_{0}$ is the density of states at the Fermi level per volume. 
Near the resonance, we can approximate 
\begin{equation}
\Gamma_{\text{qp}}^{ 20\to 11 }(\varepsilon)=\Gamma_{\text{qp}}^{ 11\to 20 } (\varepsilon) \approx\frac{1}{2\pi}g_{c} \sqrt{\frac{2 \Delta}{\pi T}} \nu_{0}^{-1} n_{\text{qp}} , \quad \varepsilon\ll T \, . \label{eq:GammaqpResonance}
\end{equation}
The precise form of $\Gamma_{\text{qp}}^{ 20\to 11 } (\varepsilon)$ near the resonance ($\varepsilon \ll T$) depends on the density of states of the superconducting island near the band edge; for example, at zero magnetic field the BCS singularity in an s-wave superconductor leads to a log-divergence,~\cite{catelani2011}  $\Gamma_{\text{qp}}^{ 20\to 11 }(\varepsilon) \propto 
 \ln \frac{2T}{|\varepsilon|}$. 
 Although we derived Eqs.~(\ref{eq:Gammaqp})--(\ref{eq:GammaqpResonance}) by using the wave functions of an s-wave superconductor, we expect the result to be qualitatively similar in the topological regime.

Finally, in Eqs.~(\ref{eq:Gammaqp})--(\ref{eq:GammaqpResonance}) we introduced the density $n_{\text{qp}}$ of thermally-activated quasiparticles, $n_{\text{qp}} =  \sqrt{\frac{\pi T\Delta}{2}} \nu_{0} e^{-\Delta/T}$~\cite{tinkham2004introduction,catelani2011}. 
In practice, the density of above-gap quasiparticles is often much larger than its equilibrium value~\cite{glazman2020bogoliubov}.

With Eq.~(\ref{eq:GammaqpResonance})  we can now estimate the smallest measurable Majorana hybridization $E_M^* \sim \Gamma_{\text{qp}}^{ 20\to 11 }(0)$ for a given quasiparticle density. 
We take estimates from   QP density in Al without taking into account QP kinetics,  which is the regime  relevant for the large volume of the  superconductor~\cite{karzig2021quasiparticle}. 
For the Al shell, we have~\cite{2015NatPh..11.1017H,PhysRevB.93.235431} a density of states per volume $\nu_0 \approx 10/(\text{eVnm}^3) $. 
At zero magnetic field and $T=20\text{mK}$, Refs.~\cite{riste2013,PhysRevLett.103.097002,PhysRevB.78.024503,PhysRevB.91.195434} give for Al a range $n_{\text{qp}}^{\text{non-eq}}= 10^{10}\text{cm}^{-3} - 10^{13}\text{cm}^{-3} $~\footnote{Density of equilibrium quasiparticles is negligible~\cite{catelani2011} even at finite field (in-field gap $\Delta=60\mu\text{eV}$) $n_{\text{qp}}^{\text{eq}} \approx 10^{2}cm^{-3}$.}. 
Taking $g_c \approx 1$, such  a range of (low to high) QP densities results in $E_M^* \approx 10^{-2} \mu \text{eV} - 10 \mu \text{eV}$. 
Thus the outlined method of measuring $E_M$ should not be limited by QP transport as long as  $E_M > E_M^*$. 
We emphasize that these estimates for $E_M^*$ are not limited by the temperature and that   accounting for the QP  kinetics would decrease $E_M^*$ even further~\cite{karzig2021quasiparticle}.  

\textbf{\emph{Current after a voltage pulse.}} 
Finally, we discuss  the possibility to measure Majorana hybridization $E_M$ in time-domain experiments. 
 Perhaps the  simplest way to study coherent dynamics of the hybridized Majorana states is by using a voltage pulse experiment~\cite{nakamura1999coherent,PhysRevX.6.031016,Tuovinen2019}. One first initializes the system in the ground state (e.g., $|20 \rangle$) by tuning $\varepsilon \rightarrow E_C \gg E_M$.   
 %for a long time $t_1 \gg 1/\Gamma_{\Sigma}$. 
 Then, at time $t=0$ one suddenly brings the system to the resonance $\varepsilon \to 0$ for a time $t_0 \sim \hbar/ E_M \ll 1/\Gamma_{\Sigma}$. 
The system evolves according to Eq.~(\ref{eq:dpdt}) and, thus, at time $t = t_0$ the population $\mathcal{P}(t_0) =p_{11;11}(t_0) + p_{21;21}(t_0)$ will be transferred to the states $|11 \rangle$ and $|21 \rangle$. 
This probability shows coherent Majorana oscillations: 
in the symmetric case and without quasiparticles~\cite{Note1} 
$\mathcal{P}(t_0) \approx \frac{I(0)}{e \Gamma_L} \left( 1 -\cos E_{M}t_0 / \hbar\right)$, where $I(0)$ is the current at $\varepsilon=0$, see Eq.~(\ref{eq:currentIdeal}).  
In order to infer $\mathcal{P}(t_0)$ from current measurement, one next detunes the system away from resonance by setting $\varepsilon \approx E_C$ for the duration $t_1 \gg 1/\Gamma_{\Sigma}$. 
In general, $\mathcal{P}(t > t_0)$ will decay slowly to the initial (ground) state with a rate controlled by $\Gamma_L$ and $\Gamma_R$ (in the absence of QP tunneling). During this decay, a current  $I(t) = e \Gamma_R \mathcal{P}(t)$ is generated, % which can be measured to infer  the population $\mathcal{P}(t_0)$, 
 see above  Eq.~(\ref{eq:I}). 
 The time-averaged current over the pulse is dominated by the decay contribution, and one finds 
%One finds time-averaged current 
 $\bar{I} \approx  e (2\Gamma_R /\Gamma_{\Sigma} ) \mathcal{P}(t_0) /t_1  $.  
Thus, current measurement gives direct access to the  coherent Majorana oscillations in $\mathcal{P}(t_0) $. Alternatively, one can measure the probability $\mathcal{P}(t_0)$ directly by charge sensing~\cite{sabonis2019dispersive}.

\textbf{\emph{Conclusions.}} 
We analyzed theoretically a way to measure the Majorana hybridization energy  $E_M$ in ac and dc transport experiments through a double-island device.
The former involves a measurement of conductance peak broadening as a function of island detuning at large bias. The latter requires measurements of charge dynamics and dephasing of $E_M$-induced coherent oscillations in the presence of a voltage pulse. %
In this work we estimated intrinsic dephasing due to above-gap QP tunneling which provides a lower bound on measurable $E_M$ value. 
We believe that our proposal can be readily implemented in both nanowire~\cite{Lutchyn_REVIEW2018} and TI~\cite{Fu09,PhysRevB.84.201105,PhysRevB.98.205403} Majorana platforms. 
Our method provides a valuable alternative to measuring $E_M$ using electromagnetic response to microwaves~\cite{2014NatCo...5E4772G,Vayrynen_2015,PhysRevB.92.245432,10.21468/SciPostPhys.7.4.050,avila2020superconducting,avila2020majorana}.

{\bf \emph{Acknowledgments.}} We thank Charlie Marcus for stimulating discussions; JIV thanks Leonid  Glazman and Bernard van Heck for discussions and  previous collaboration on a related project. This work is based on support by the U.S. Department of Energy, Office of Science through the Quantum Science Center (QSC), a National Quantum Information Science Research Center.

\bibliographystyle{apsrev4-1}
\bibliography{refs}

%\newpage

% The following merges the supplement into the main text


\section*{Supplementary material}

# SUPPLEMENTARY MATERIAL TO "MAJORANA SIGNATURES IN CHARGE TRANSPORT THROUGH A TOPOLOGICAL SUPERCONDUCTING DOUBLE-ISLAND SYSTEM"

In this Supplementary Material, we present the detailed electrostatic Hamiltonian, the Bogoliubov transformation leading to Eqs. (5)–(6) of the main text, details for the master equation and average current, and numerical simulation and an analytical result for the coherent oscillations.

## I. CHARGING ENERGIES IN TERMS OF CIRCUIT CAPACITANCES.

Let us denote by $C_{gL,R}, C_{L,R}, C_m$ the islands' capacitances to the gates, leads, and their mutual capacitance. Ignoring cross-capacitances, the total capacitances of the two islands are $C_{1,2} = C_{gL,R} + C_{L,R} + C_m$. The charging energies and induced charges in Eq. (2) of the main text can be related to junction capacitances and gate charges [1]:

$$E_C^{L,R} = \frac{1}{2} e^2 \frac{C_{2,1}}{C_1 C_2 - C_m^2}, \quad E_C^m = e^2 \frac{C_m}{C_1 C_2 - C_m^2}. \tag{1}$$

Since $C_1 + C_2 > 2C_m$, we have $E_C^L + E_C^R > E_C^m$. The induced charges are

$$N_{gL} = \frac{1}{|e|} \left\{ C_{gL} V_{gL} + C_L V_L + (C_{gR} V_{gR} + C_R V_R) \frac{E_C^m}{2 E_C^L} \right\}, \tag{2}$$

$$N_{gR} = \frac{1}{|e|} \left\{ C_{gR} V_{gR} + C_R V_R + (C_{gL} V_{gL} + C_L V_L) \frac{E_C^m}{2 E_C^R} \right\}. \tag{3}$$

It is possible to introduce gauge-invariant gate charges,

$$\mathcal{N}_{gL} = \frac{1}{|e|} \left\{ C_{gL}(V_{gL} - V_L) + [C_{gR}(V_{gR} - V_L) + C_R(V_R - V_L)] \frac{E_C^m}{2 E_C^L} \right\}, \tag{4}$$

$$\mathcal{N}_{gR} = \frac{1}{|e|} \left\{ C_{gR}(V_{gR} - V_L) + C_R(V_R - V_L) + C_{gL}(V_{gL} - V_L) \frac{E_C^m}{2 E_C^R} \right\} \tag{5}$$

which satisfy

$$2 E_C^L \mathcal{N}_{gL} = 2 E_C^L N_{gL} + |e| V_L \tag{6}$$
$$2 E_C^R \mathcal{N}_{gR} = 2 E_C^R N_{gR} + |e| V_L \tag{7}$$

In this choice of gauge, every voltage is measured relative to the left lead. It is then convenient to take the left lead to be grounded, $V_L \to 0$.

## II. BOGOLIUBOV TRANSFORMATION LEADING TO EQS. (5)–(6).

Equations (5)–(6) of the main text can be derived by considering the following phenomenological model that captures the essential physics. We start by considering a superconducting island with s-wave pairing, ($\tau = \pm$ and $n$ denote the spin and orbital indices)

$$H_{\text{island}} = \sum_{n,\tau} \xi_n c_{n,\tau}^\dagger c_{n,\tau} + \Delta \sum_n (e^{-i\varphi} c_{n,-} c_{n,+} + e^{i\varphi} c_{n,+}^\dagger c_{n,-}^\dagger), \tag{8}$$

where $e^{i\varphi}$ decreases the charge of the condensate by 2, $e^{i\varphi} N^c = e^{i\varphi}(N^c - 2)$ so that the model is charge conserving. The total charge operator of the island is $N = N^c + \sum_{n,\tau} c_{n,\tau}^\dagger c_{n,\tau}$. We consider here a single island and leave out the index $\alpha = L, R$ used in the double-island case. Even though $H_{\text{island}}$ is time-reversal symmetric, we do not expect qualitative changes in the presence of broken time-reversal symmetry, see remarks below Eq. (12). The Hamiltonian $H_{\text{island}}$ can be diagonalized with a Bogoliubov transformation

$$c_{n,\tau} = e^{i\varphi/2}(d_{n,\tau} u_n - \tau v_n^* d_{n,-\tau}^\dagger), \quad d_{n,\tau} = u_n^* c_{n,\tau} e^{-i\varphi/2} + \tau v_n^* c_{n,-\tau}^\dagger e^{i\varphi/2}. \tag{9}$$

The operators $d_{n,\tau}$ correspond to neutral quasiparticles since they commute with $N$. The coherence factors are

$$u_n = \sqrt{\frac{1}{2}\left(1 + \frac{\xi_n}{\sqrt{\xi_n^2 + |\Delta|^2}}\right)}, \quad v_n = \sqrt{\frac{1}{2}\left(1 - \frac{\xi_n}{\sqrt{\xi_n^2 + |\Delta|^2}}\right)}. \tag{10}$$

The above diagonalization for both the left and the right islands leads to the Hamiltonian (5) of the main text. In the main text we dispensed with the spin index $\tau$ which is not expected to be a good quantum number in the realistic case with spin-orbit coupling and broken time-reversal symmetry. In our model, the Majorana operators $\gamma, \gamma'$ are included phenomenologically by including a zero-mode $d = (\gamma - i\gamma')/2$ in the diagonalized Hamiltonian.

The tunneling of electrons between the islands can be described with the Hamiltonian

$$H_{LR} = \sum_{nn'\tau} t_{nn'} c^\dagger_{n,\tau,L} c_{n',\tau,R} + h.c. \tag{11}$$

$$= \sum_{nn'\tau} t_{nn'} N_L^+ N_R^- (d^\dagger_{n,\tau,L} u^*_{nL} - \tau v_{nL} d_{n,-\tau,L})(d_{n',\tau,R} u_{n'R} - \tau v^*_{n'R} d^\dagger_{n',-\tau,R}) + h.c., \tag{12}$$

where we introduced the labels for the islands and denote $e^{i\varphi_{L,R}/2} = N_{L,R}^- = (N_{L,R}^+)^\dagger$. In the main text, Eq. (??), we have also included phenomenologically the hybridization term $\propto E_M$ between the inner Majorana modes. Finally, we note that $d^\dagger_{n,\tau,\alpha}$ creates a quasiparticle with energy $\varepsilon_{n,\alpha} = \sqrt{\xi_{n,\alpha}^2 + |\Delta|^2} > \Delta$. Thus, the terms such as $\sim d^\dagger_{n,\tau,L} d^\dagger_{n',-\tau,R}$ in the tunneling correspond to a pair-breaking excitation with energy $\gtrsim 2\Delta$ [or $\gtrsim \Delta$ if one Majorana level is involved]. Assuming low bias voltage and temperature, these terms are not allowed in the sequential tunneling and were dropped from Eq. (??) of the main text.

### III. GENERAL EVOLUTION OF THE DENSITY MATRIX.

In Eq. (??) of the main text we took $n_L(\varepsilon_L^{(\prime)}) = 1 - n_R(\varepsilon_R^{(\prime)}) = 1$ valid at high bias voltage. Without this assumption, the full form of Eq. (??) is

$$\frac{d}{dt} p_{20;20} = -E_M \operatorname{Im} p_{20;11} + \Gamma_{qp}^{11\to 20} p_{11;11} - \Gamma_{qp}^{20\to 11} p_{20;20} \tag{13}$$

$$+ \Gamma_L p_{10;10} n_L(\varepsilon_L) - \Gamma_L p_{20;20}[1 - n_L(\varepsilon_L)] - \Gamma_R p_{20;20} n_R(\varepsilon_R) + \Gamma_R p_{21;21}[1 - n_R(\varepsilon_R)] \tag{14}$$

$$\frac{d}{dt} p_{11;11} = E_M \operatorname{Im} p_{20;11} - \Gamma_{qp}^{11\to 20} p_{11;11} + \Gamma_{qp}^{20\to 11} p_{20;20} \tag{15}$$

$$+ \Gamma_L p_{21;21}[1 - n_L(\varepsilon'_L)] - \Gamma_L p_{11;11} n_L(\varepsilon'_L) - \Gamma_R p_{11;11}[1 - n_R(\varepsilon'_R)] + \Gamma_R p_{10;10} n_R(\varepsilon'_R) \tag{16}$$

$$\frac{d}{dt} p_{10;10} = \Gamma_R p_{11;11}[1 - n_R(\varepsilon'_R)] + \Gamma_L p_{20;20}[1 - n_L(\varepsilon_L)] - \Gamma_L p_{10;10} n_L(\varepsilon_L) - \Gamma_R p_{10;10} n_R(\varepsilon'_R) \tag{17}$$

$$\frac{d}{dt} p_{21;21} = \Gamma_L p_{11;11} n_L(\varepsilon'_L) + \Gamma_R p_{20;20} n_R(\varepsilon_R) - \Gamma_R p_{21;21}[1 - n_R(\varepsilon_R)] - \Gamma_L p_{21;21}[1 - n_L(\varepsilon'_L)] \tag{18}$$

$$\frac{d}{dt} p_{20;11} = i\varepsilon p_{20;11} - i\frac{1}{2} E_M (p_{11;11} - p_{20;20}) - \frac{1}{2} \Gamma_\Sigma p_{20;11} \tag{19}$$

where $n_s(\epsilon) = (e^{(\epsilon - \mu_s)/T} + 1)^{-1}$ is the Fermi distribution function of the $s = L, R$ lead. The high-bias limit $n_L(\varepsilon_L^{(\prime)}) = 1 - n_R(\varepsilon_R^{(\prime)}) = 1$ requires $eV_b \equiv |\mu_L - \mu_R| > E_C^m$ under symmetric biasing at the resonance $\varepsilon \approx 0$ and $N_{g,tot} = 2$. Here $N_{g,tot} = 2\frac{E_C^L N_{gL} + E_C^R N_{gR}}{E_{C,tot}}$ is the "center of mass" dimensionless gate charge and $E_{C,tot} = E_C^R + E_C^L + E_C^m$ is the total charging energy. The more general condition on the lower bound for the bias voltage depends on both $N_{g,tot}$ and $N_g = \frac{E_C^R N_{gR} - E_C^L N_{gL}}{E_C}$.

The upper bound on bias voltage is set by the condition that above-gap quasiparticles are not excited. In the simple case of an s-wave superconductor with a sharp density of states, this condition can be written as $eV_b < \Delta - T$ where $\Delta$ is the s-wave gap.

## IV. THE AVERAGE CURRENT.

The current operator can be taken to be the rate of change of the number of electrons in the right lead, $\hat{I} = e\partial_t \sum_{k'} c^\dagger_{k'R} c_{k'R}$. From the Heisenberg equation of motion we find,

$$e\langle \frac{d}{dt} \sum_{k'} c^\dagger_{k'R} c_{k'R} \rangle = ie\langle [H, \sum_{k'} c^\dagger_{k'R} c_{k'R}]\rangle \tag{20}$$

$$= ie \sum_{k'k} \left\langle t_R N_R^+ [\gamma_R c_{kR}, c^\dagger_{k'R} c_{k'R}] \right\rangle + h.c. \tag{21}$$

$$= 2e \text{Im}\, t_R \sum_k \left\langle c_{kR} \gamma_R (\hat{P}_{21;20} + \hat{P}_{11;10}) \right\rangle, \tag{22}$$

where we projected to the relevant charge states. We use the Kubo formula to calculate the averages,

$$\text{Im}\, t_R \left\langle c_{kR} \gamma_R \hat{P}_{21;20} \right\rangle (t) = -\text{Im}\, it_R^* \int_{-\infty}^t dt' e^{0t'} e^{-i(\varepsilon_{kR} - \varepsilon_R)(t-t')} \left\langle c^\dagger_{kR} c_{kR} \hat{P}_{20;20} - c_{kR} c^\dagger_{kR} \hat{P}_{21;21} \right\rangle \tag{23}$$

$$\approx -\pi |t_R|^2 \delta(\varepsilon_{kR} - \varepsilon_R) \left[p_{20;20} n_{kR} - p_{21;21}(1 - n_{kR})\right], \tag{24}$$

where we factored the average and approximated the slowly-evolving island averages by their static steady-state values. Similarly, we find

$$\text{Im}\, t_R \left\langle c_{kR} \gamma_R \hat{P}_{11;10} \right\rangle (t) = -\pi |t_R|^2 \delta(\xi_{kR} - \varepsilon'_R) \left[n_{kR} p_{10;10} - (1 - n_{kR}) p_{11;11}\right]. \tag{25}$$

We find therefore the current used to derive Eq. (9) of the main text,

$$\langle \hat{I} \rangle = e\Gamma_R \{p_{21;21}[1 - n_R(\varepsilon_R)] + p_{11;11}[1 - n_R(\varepsilon'_R)] - p_{20;20} n_R(\varepsilon_R) - p_{10;10} n_R(\varepsilon'_R)\}. \tag{26}$$

## V. COHERENT OSCILLATIONS AFTER A VOLTAGE PULSE.

We can use Eq. (8) of the main text to also study charge dynamics. We consider a pulse shape that is a piecewise function, with $\varepsilon(t) = \varepsilon_0$ for $0 < t < t_0$ and $\varepsilon(t) = E_C$ for $t_0 < t < t_0 + t_1$. We take $t_0 \sim 1/E_M \ll 1/\Gamma_\Sigma \ll t_1$. Mathematically, a suitable function of period $t_0 + t_1$ that does this is $\varepsilon(t) \approx \varepsilon_0 + (E_C - \varepsilon_0)\Theta(t - (t_0 + t_1)\lfloor \frac{t}{t_0+t_1} \rfloor - t_0)$ where $\Theta$ and $\lfloor . \rfloor$ denote the step and floor functions. We denote by $\varepsilon_0 \ll E_C$ the gate-charge value near resonance; in the main text we consider a pulse exactly into resonance, $\varepsilon_0 = 0$. A numerical simulation for the probability $\mathcal{P}(t_0, \varepsilon_0) = p_{11;11}(t_0) + p_{21;21}(t_0)$ is shown in Fig. 1, with (a) weak and (b) strong quasiparticle tunneling rates. $\mathcal{P}(t_0)$ shows coherent oscillations at frequency $E_M$ which enables the dynamical observation of the Majorana hybridization.

A simple analytical expression for $\mathcal{P}(t, 0)$, and its coherent oscillations, can be obtained in the symmetric case ($\Gamma_R = \Gamma_L$) and in absence of quasiparticles ($\Gamma_\Sigma/2 = \Gamma_L$): by taking initial condition $p_{20;20}(0) = 1$ in Eq. (8) of the main text, we find $\mathcal{P}(t_0, 0) \approx \frac{I(0)}{e\Gamma_L}(1 - \cos E_M t_0/\hbar)$, where $I(0)$ is the current at resonance ($\varepsilon = 0$), see Eq. (1) of the main text. For $t > t_0$, we have $\varepsilon = E_C$ and the population $\mathcal{P}(t, 0)$ will decay exponentially (as charge tunnels incoherently into the lead). During this decay a current is generated (see text above Eq. (9) of the main text). We evaluate the time-averaged current, $\bar{I} = \frac{1}{t_0+t_1}[\int_0^{t_0} dt I(t) + \int_{t_0}^{t_1+t_0} dt I(t)]$. It is dominated by the second term since $t_0 \ll t_1$, and we thus have $\bar{I} \approx e(2\Gamma_R/\Gamma_\Sigma)\mathcal{P}(t_0)/t_1$ with $\Gamma_\Sigma \approx \Gamma_R + \Gamma_L$. Here we neglected subleading contributions proportional to $E_M/E_C \ll 1$ and $\Gamma_\Sigma t_0 \ll 1$. Thus, the averaged current $\bar{I}$ over the pulse duration oscillates in $t_0$ with frequency $E_M$.

---

[1] W. G. van der Wiel, S. De Franceschi, J. M. Elzerman, T. Fujisawa, S. Tarucha, and L. P. Kouwenhoven, Rev. Mod. Phys. **75**, 1 (2002).

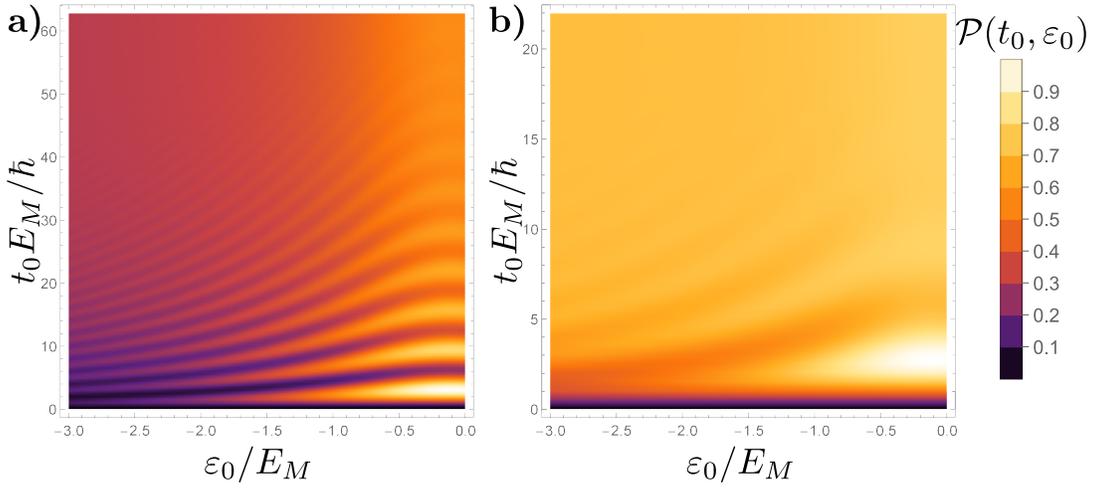

Figure 1. Coherent oscillations in the probability $\mathcal{P}(t_0, \varepsilon_0) = p_{11;11}(t_0, \varepsilon_0) + p_{21;21}(t_0, \varepsilon_0)$ following a pulse in gate charge, $\varepsilon \to \varepsilon_0$, in the cases of **a)** weak $[\Gamma_{\text{qp}}^{20 \to 11} = 0.5 \Gamma_L]$ and **b)** strong $[\Gamma_{\text{qp}}^{20 \to 11} = 6 \Gamma_L]$ quasiparticle tunneling. In the the text we denoted $\mathcal{P}(t_0) \equiv \mathcal{P}(t_0, 0)$, discussing the case $\varepsilon_0 = 0$. It is assumed that the state before the pulse is initialized in $|20\rangle$. We have also taken the large-bias limit $n_L = 1 - n_R = 1$ and assumed $\varepsilon_0 \ll T$ so that $\Gamma_{\text{qp}}^{20 \to 11} = \Gamma_{\text{qp}}^{11 \to 20}$. In both plots we have $\Gamma_L = \Gamma_R = 0.05 E_M$ and $\Gamma = 0$.



\end{document}